# Application of the Gratton Vargas Model to arbitrary anode and insulator shapes

S K H Auluck
International Scientific Committee for Dense Magnetized Plasmas,
http://www.icdmp.pl/isc-dmp
Hery 23, P.O. Box 49, 00-908 Warsaw, Poland

## Abstract

This paper demonstrates the application of the Gratton-Vargas model (GV) 2-D snowplow model to arbitrary anode and insulator geometry. This is an abridged version of a larger project of constructing an incremental model of the plasma focus and similar devices that mainly addresses the transfer of energy from the energy storage to the plasma until it reaches the axis since the physics of plasma transformations that follow is still at the experimental discovery stage. The first stage of this incremental model is to construct a kinematic framework that allows calculation of an imaginary 3-D surface of rotation that serves the same purpose as the imaginary center of mass in mechanics. The procedure is illustrated with a simple anode shape.

## I. Introduction:

The Gratton-Vargas (GV) model[1] of the Dense Plasma Focus (DPF) [2,3] recently re-visited and revised[4-6] has demonstrated its ability [7] to fit experimental current waveforms from multiple installations, its utility as a scaling theory of device optimization [8,9] and its potential for providing a framework for constructing first-principles analyses of the DPF physics based on Hyperbolic Conservation Law (HCL) formalism[10,11,12]. This paper demonstrates its application to arbitrary anode an insulator shapes. This is an abridged version of a larger project that aims to construct an incremental model of the plasma focus and similar devices that mainly addresses the transfer of energy from the energy storage to the plasma until it reaches the axis since the physics of plasma transformations that follow is still at the experimental discovery stage.

A unique aspect of the GV model is its kinematic and analytical nature. The GV model derives its results from mathematical properties of the analytical solution to a partial differential equation resulting from an ad hoc hypothesis and not from well-established fundamental principles of physics such as laws of mechanics and electrodynamics or physical phenomena such as heat transport, electrical resistivity, atomic and molecular processes involving collisions / excitation / dissociation / ionization / radiation etc. There is no reason, rooted in physics, for this kind of model to yield results that compare well with experiments. *That this model still manages to reproduce current waveforms[7] and plasma shapes similar to those experimentally determined is itself a significant physical datum, that retrospectively justifies the ad hoc assumptions made while deriving the partial differential equation.*

A physical theory of plasma, by its very nature, is approximate: it is based on simplifying assumptions that assert that certain attributes of the plasma play a negligible role in governing some of its properties. A hierarchy of theories, which progressively relax a corresponding hierarchy of simplifying assumptions, is necessary to construct an 'asymptotically complete' description of plasma. A kinematic model, which represents the shape and position of the plasma by an imaginary 3-dimensional surface (as a cartoon represents a human figure), may be considered as the ground level of such hierarchy of theories, which neglects almost all physical processes except perhaps some rudimentary description of "motion" in a generalized sense. By comparing predictions of such kinematic model with experimental observations, one can determine that many important physical phenomena possibly play a negligible role in determining some aspects of plasma behavior. This is an important distinction. For example, it has been argued [10] that the temporal current profile of the DPF is insensitive to the details of transport of heat and electricity, radiative, atomic and molecular phenomena, properties of electrode and insulator materials (which have all been taken into consideration in a 3D simulation[15]) or of conservation laws. It should be important to investigate whether there are other attributes of the DPF which are similarly insensitive.

The GV model is based on the *a priori* assumption of equality between the magnetic pressure due to the azimuthal magnetic field behind the plasma current sheath (PCS) and the density times the square of the normal component of sheath velocity. This could be "justified" as a boundary condition on the propagating PCS. Sufficiently far behind it, the magnetic pressure is the only force acting on a unit area of the PCS and sufficiently far ahead, the neutral gas of mass density $\rho_0$ getting ingested by the PCS moving with a normal velocity $v_{PCS}$ brings in a momentum density $-\rho_0 v_{PCS}$ with a momentum density flux $(-\rho_0 v_{PCS})(-v_{PCS})$. Newton's second law then suggests equality of the force per unit area with the momentum density flux. *This however is not a rigorous application of Newton's second law* and must be treated as a model assumption. There could be other ways of "justifying" this equality but they all involve some assumptions that have no basis either in theory or in experiments. This model assumption happens to be supported by the 2-D MHD simulations of Maxon and Eddleman [16] and experimental data[17].

This equality helps in defining a dimensionless time-equivalent parameter $\tau$, which equals the charge flow in real time t normalized to a 'mechanical equivalent of charge'[1] – a quantity having dimensions of charge constructed from physical parameters of the DPF. This equality also enables the GV model to construct a partial differential equation for the time and space evolution of an imaginary 3-dimensional surface of revolution – called GV surface – that is the anchor of this equality. It turns out that this family of 3-D surfaces resembles the experimentally observed shape of the PCS[17 ].

Assuming that this family of GV surfaces approximates the actual path of current flow, its inductance can be calculated as a function of the dimensionless time-equivalent parameter $\tau$. The circuit equation then provides the current profile as a function of $\tau$, from which, the corresponding real time t can be determined as the *last procedure* of the model. *This ability to decouple the circuit equation from the physical motion of the plasma in real time is a feature that distinguishes the GV model from all other theories and models of DPF*. The original work of Gratton and Vargas [1] was unable to take into account the circuit resistance, without which, it was not possible to compare the model with experimental current waveforms. This was accomplished more recently[4] leading to the Resistive Gratton-Vargas (RGV) model.

The motivation for extending the discussion to non-standard adaptations of the simple Mather geometry arises from the fact that many laboratories do work with anodes which are tubular or have a cavity or protrusion near the axis or have a hemispherical or spherical anode. Presently, there is no theoretical framework for comparing the effect of such deviations on the operation taking the simple geometry as a reference. This also may find applications in the design of plasma flow switches that transfer energy from the energy store to the load efficiently and with a shorter rise time.

The present paper revisits the construction of the GV model to give a concise and coherent review of the work referred above in a single logically consistent narrative and along the way shows how it is applied to an arbitrary geometry. While earlier work[4,5], was more in the nature of tutorial introduction to the nearly forgotten work of Gratton and Vargas[1] after a gap of several decades, the emphasis in this paper is on a comprehensive presentation of the GV model (including some missing details).

This paper is organized as follows. Section II derives a generalization of the RGV model that is valid for generalized analytically defined shapes of anode and insulator. An illustration of the method for complex anode and insulator shape is presented in Section III. Section IV summarizes the paper.

## II. Generalization of RGV model including modified geometry, sheath propagation delay

The RGV model [4,5] assumes that the DPF consists of a straight cylindrical anode of radius 'a' and length $\mathbb{L}_a$, and insulator of outer radius $\mathbb{R}_I$ and length $\mathbb{L}_I$ and a cathode of inner radius $\mathbb{R}_C$; is filled with a gas having mass density $\rho_0$ and is powered by a capacitor bank that is idealized by a single capacitance $C_0$, in series with a constant resistance $R_0$ , a constant inductance $L_0$ and an ideal closing switch (including its connection hardware), whose contribution to circuit inductance, circuit resistance and circuit capacitance is assumed to be included in $L_0$, $R_0$ and $C_0$.

*This idealization of circuit may not be valid in some DPF devices*. In some small devices, the switch resistance may have significant variation as compared with the circuit impedance over a significant portion of the quarter cycle time. In some large devices, multiple spark gaps may begin conduction at significantly different times. In both cases, the assumed circuit model is not expected to be valid. This needs to be kept in mind when a fitting procedure fails to yield a good fit to the experimental current waveform with the RGV model or the present model.

In this section, this description of the DPF is relaxed to include arbitrary shapes of anode and insulator, which are represented by surfaces of revolution about the z-axis in cylindrical coordinates. *Not all shapes of anode and insulator are amenable to this treatment – the limitations are pointed out later in the discussion*. For this purpose, 'a' is re-defined as a normalizing scale length related to the anode size. The anode and insulator shapes are defined in terms of a coordinate $\tilde{Z}$ (which is used only for specifying boundaries) measured in units of scale length 'a' along the cylindrical axis starting from the base of the anode. In a two dimensional representation of these solid surfaces in $(r,z)$ space, the anode would be a curve that has a finite radius at its base $\tilde{Z}=0$ and would have a maximum height $\tilde{Z}=\tilde{z}_A$, the normalized length of the anode. Similarly, the insulator would be a curve that begins at $\tilde{Z}=0$ with a finite radius and *meets the anode* at $\tilde{Z}=\tilde{z}_I$, the normalized length of the insulator. The normalized anode radius $\tilde{R}_A$ and normalized insulator radius $\tilde{R}_I$ are specified as functions $\tilde{R}_A(\tilde{Z})$ and $\tilde{R}_I(\tilde{Z})$ of $\tilde{Z}$, which may be piecewise continuous with arbitrary number of segments:

$$\tilde{R}_A\left(\tilde{Z}\right)=\tilde{R}^{(i)}_{A,ext}\left(\tilde{Z}\right),\quad \tilde{Z}^{(i-1)}_A\le\tilde{Z}\le\tilde{Z}^{(i)}_A, i=1,2\cdots,i_m;\quad \tilde{R}^{(i)}_{A,ext}\left(\tilde{Z}^{(i)}_A\right)=\tilde{R}^{(i-1)}_{A,ext}\left(\tilde{Z}^{(i)}_A\right);$$
$$=\tilde{R}^{(j)}_{A,int}\left(\tilde{Z}\right),\quad \tilde{Z}^{(i_m+j-1)}_A\ge\tilde{Z}\ge\tilde{Z}^{(i_m+j)}_A, j=1,2\cdots,j_m;\quad \tilde{R}^{(j)}_{A,int}\left(\tilde{Z}^{(i_m+j)}_A\right)=\tilde{R}^{(j+1)}_{A,int}\left(\tilde{Z}^{(i_m+j)}_A\right); \tag{1}$$

$$\tilde{R}_I\left(\tilde{Z}\right)=\tilde{R}^{(i)}_I\left(\tilde{Z}\right),\quad \tilde{Z}^{(i-1)}_I\le\tilde{Z}\le\tilde{Z}^{(i)}_I, i=1,2\cdots,i_n;\quad \tilde{R}^{(i)}_I\left(\tilde{Z}^{(i)}_I\right)=\tilde{R}^{(i-1)}_I\left(\tilde{Z}^{(i)}_I\right); \tag{2}$$

The anode profile consists of two branches: an external branch consisting of segments $\tilde{R}^{(i)}_{A,ext}\left(\tilde{Z}\right)$ , where the coordinate $\tilde{Z}$ increases monotonically till it reaches a maximum value $\tilde{Z}^{(i_m)}_A$ , which is designated as the normalized anode length $\tilde{z}_A$. After that, the anode profile may or may not have an internal branch representing a cavity, consisting of segments $\tilde{R}^{(j)}_{A,int}\left(\tilde{Z}\right)$, where $\tilde{Z}$ would decrease monotonically. The anode profile segment functions $\tilde{R}^{(i)}_{A,ext}\left(\tilde{Z}\right)$ and $\tilde{R}^{(j)}_{A,int}\left(\tilde{Z}\right)$ will not henceforth be distinguished in the text unless the context requires such distinction and both will be referred as $\tilde{R}^{(i)}_A\left(\tilde{Z}\right)$. The points $\left(\tilde{R}^{(i)}_A\left(\tilde{Z}^{(i)}_A\right),\tilde{Z}^{(i)}_A\right)$ etc. shall be referred as vertices. The following convention is also adopted: $\tilde{Z}^{(0)}_A\equiv\tilde{z}_I$, the normalized length of the insulator and $\tilde{Z}^{(0)}_I=0$.

The cathode is still taken to be a hollow cylinder with inner normalized radius $\tilde{r}_C$ connected to a metal base plate lying in the $\tilde{Z}=0$ plane. In this notation, the classical Mather type plasma focus considered in earlier work[4,5] on the RGV model is described by the functions

$$\begin{aligned}\tilde{R}_A\left(\tilde{Z}\right)&=1, && 0\le\tilde{Z}\le\tilde{z}_A,\\ &=r && 0\le r\le 1,\tilde{Z}=\tilde{z}_A\\ \tilde{R}_I\left(\tilde{Z}\right)&=\tilde{r}_I>1, && 0\le\tilde{Z}\le\tilde{z}_I\\ &=r && 1\le r\le\tilde{r}_I,\tilde{Z}=\tilde{z}_I\end{aligned} \tag{3}$$

An illustration of the present model with a complex anode profile is discussed in Section III.

After the (ideal) switch operates at time t=0, the discharge plasma remains stationary for some time $t_s$ and then begins to move. In this abridged version, this propagation delay $t_s$ is neglected.

Then the plasma begins to expand both because of its thermal pressure that builds up on account of resistive heating and the magnetic pressure that pushes the current carrying plasma away from the current carrying anode[13]. Since the speed of expansion of the plasma exceeds the thermal velocity in the neutral gas, a hydrodynamic shock wave is established, whose speed depends on the continually increasing driving pressure that is predominantly determined by the magnetic pressure exerted on the current carrying plasma. The hydrodynamic shock wave becomes an ionizing shock wave[19] when the photons produced in the plasma travel ahead of the shock wave, get absorbed in the neutral gas layer adjacent to the shock wave and produce photoelectrons. The current carrying layer of plasma has an electric field[19] in a direction perpendicular to the direction of propagation of plasma that is of the order of $\vec{v} \times \vec{B}$ where $\vec{v}$ is the fluid velocity of the plasma and $\vec{B}$ is the magnetic field in the current carrying layer. Since transverse component of electric field is continuous across interfaces between dissimilar media, this electric field extends into the neutral gas layer adjacent to the hydrodynamic shock wave, accelerates the photo-electrons, dissociates and ionizes the neutral gas layer by electron impact ionization and further heats it resistively[19]. As a result the adjacent layer of neutral gas gets incorporated in the moving ionizing shock wave, which eventually evolves into an autonomous structure consisting of a fully ionized plasma zone that carries the bulk of the current and another transition zone [14] that contains a partially ionized plasma which is bounded on one side by neutral gas and on the other side by a fully ionized plasma, capable of propagating by itself into the neutral gas. This autonomous structure shall be referred as a plasma current sheath (PCS) and its transition zone between the neutral gas and the fully ionized plasma shall be called as the dense plasma sheath (DPS). This scenario is often referred as the snowplow model since it assumes that the neutral gas layer adjacent to the hydrodynamic shock wave gets fully incorporated in the PCS just as the snowplow gathers the entire layer of snow in its path.

This complex phenomenology, however, is subject to a "boundary condition". Sufficiently far behind the PCS, the force that is acting on a unit area is the magnetic pressure due to the current that is flowing through the anode and the plasma, which is given by $B_\theta{}^2 / 2\mu_0$. The neutral gas of mass density $\rho_0$ that gets ingested by the PCS moving with normal velocity $v_n$ brings with it a momentum density $\left(-\rho_0 v_n\right)$ so that the rate of change of momentum per unit area per unit mass is the momentum flux density $\left(-\rho_0 v_n\right)\left(-v_n\right)$. Newton's Second Law then suggests that these two quantities should be equal to ensure global momentum balance. *Note that this isn't local momentum balance: the neutral gas and the magnetic field exist in mutually exclusive regions*[20].

One can now define an imaginary surface, that is co-located with the PCS in an as-yet-undetermined manner, where the equality

$$\rho_0 {v_n}^2 = {B_\theta}^2 / 2\mu_0 \tag{4}$$

holds *by definition*. Let this surface, called the GV surface, having full azimuthal symmetry, be described in cylindrical coordinates by the equation

$$\psi\left(r,z,t\right) \equiv z - f\left(r,t\right) = 0 \tag{5}$$

By definition, the convective derivative of $\psi$ is zero:

$$\partial_t \psi + \vec{v} \cdot \vec{\nabla} \psi = 0 \tag{6}$$

where $\vec{v}$ is the velocity of the surface. The component of velocity normal to the surface is

$$v_n = \vec{v}.\left(\vec{\nabla}\psi \middle/ \left|\vec{\nabla}\psi\right|\right) = -\partial_t \psi \middle/ \sqrt{\left(\partial_r \psi\right)^2 + \left(\partial_z \psi\right)^2} \tag{7}$$

From (4) and (7), using the standard result from Ampere's Law

$$B_\theta = \frac{\mu_0 I\left(t\right)}{2\pi r} \tag{8}$$

one can write

$$\partial_t \psi + \sqrt{\left(\partial_r \psi\right)^2 + \left(\partial_z \psi\right)^2}\, \frac{\chi \mu_0 I\left(t\right)}{2\pi r \sqrt{2\rho_0 \mu_0}} = 0 \tag{9}$$

*Note that this result applies only when the PCS structure has been formed and has started propagation until an ionizing shock wave has truly been formed.*

Define now a time-equivalent dimensionless parameter $\tau$ by the equation[1,4]

$$\tau \equiv \frac{1}{Q_m}\int_0^t I(t)dt \text{ where } Q_m = \mu_0^{-1}\pi a^2\sqrt{2\mu_0\rho_0} \ , \tag{10}$$

and dimensionless coordinates $\tilde{r} \equiv r/a$, $\tilde{z} \equiv z/a$. This enables reduction of (9) to a dimensionless form

$$\partial_\tau\psi + \sqrt{\left(\partial_{\tilde{r}}\psi\right)^2 + \left(\partial_{\tilde{z}}\psi\right)^2}\,\frac{1}{2\tilde{r}} = 0 \tag{11}$$

This is the Gratton-Vargas[1,4] equation. Note that the parameter 'a', which signified anode radius in previous work[4-6], has been re-defined as a scale length related to the generalized anode shape, since the anode radius varies along its length.

This equation is based on precisely four inputs: (1) definition of normal velocity of an arbitrary propagating surface of revolution in cylindrical geometry (equation (7)) (2) the so called snowplow hypothesis (equation (4)) (3) a standard result from Ampere's Law (equation (8)) and (4) definition of the variable $\tau$ (equation (10)) . *It specifically does not involve the following*: (1) the assumption of a specific shape or size of anode, cathode or insulator or any relations between them, (2) the precise definition of the scale length 'a' which is *not* the anode radius as in earlier work[4,5], (3) the assumption that the driving circuit is a simple capacitor discharge circuit with constant intrinsic inductance and resistance.

For its application to the generalized plasma focus problem stated above, equation (11) must be considered along with the following supplementary conditions.

1. At $\tau = 0$, the initial shape of the solution corresponds to the insulator profile, on which the initial plasma is formed: $\psi(\tilde{r},\tilde{z},\tau=0) = \psi\left(\tilde{R}_I\left(\tilde{Z}\right),\tilde{Z},\tau=0\right)$
2. At every value of $\tau > 0$, the solution has a curve of intersection with the anode: $\psi\left(\tilde{R}_A\left(\tilde{Z}\right),\tilde{Z},\tau\right)$ is part of the solution of the equation at $\tau$.
3. It is assumed that the current that crosses over from the anode to the plasma does so *at or behind the GV surface.* That means that the current density at the anode surface has a zero in-surface component at the junction between the GV surface and the anode. Assuming further (in view of the similarity of experimentally determined sheath shape with the solutions of the GV equation [16]), that the GV surface approximates the shape of the path that the current density takes in the plasma, the condition of continuity of current density at the junction implies that *the GV surface must be perpendicular to the anode at the junction*. This condition applies to the forward motion of the GV surface.

4. The solution must join the anode profile with the cathode.

It is important to keep in mind that this equation is a *mere mathematical construct* having no inherent *physical* meaning apart from the assumptions explicitly mentioned. It acquires any physical meaning only when (and to the extent) it is used as a tool for understanding a physical phenomenon. For example, it cannot be taken for granted that a solution of the GV equation along with supplementary conditions will actually correspond to an experimentally realizable system. Equally, every experimentally realizable system that conforms to the problem statement may not have a representation in terms of solutions of the GV equation. The point of the exercise is that there is a *strong likelihood* of a correspondence between the GV surface calculated for an assumed device geometry and the PCS of its experimental realization. If (and only if) such correspondence is experimentally established, it becomes a *new independent datum* which a physical first-principles theory must explain or be consistent with. Equally, demonstration of absence of such correspondence in particular cases should provide important insights regarding additional conditions that may be necessary for such correspondence to exist.

The full extent of mathematical properties of this equation, the supplementary conditions and its solutions has not been explored till date and many uncertainties and doubts remain about the existence, uniqueness and character of its solution for various kinds of boundaries. A tutorial discussion on some types of solution of this equation given by Gratton and Vargas is found elsewhere [1,4,5]. Only a few salient features of the solution[1,4] using the method of characteristics[21] are brought out here.

The method of characteristics finds[21] a family of characteristic line elements (CLE) *in the neighborhood* of every point on the integral surface of the partial differential equation (1) along which its solution is constant (2) and which is locally perpendicular to the integral surface. A collection of such CLEs is referred as "characteristic curves". Since the integral surface moves with time, the characteristic curves are implicitly tagged with time, although the algebraic definition of the characteristic curve may not have time as a variable. This subtle distinction has important consequences which are discussed later.

Define generalized momenta $p_{\tilde{r}} \equiv \partial_{\tilde{r}}\psi$, $p_{\tilde{z}} \equiv \partial_{\tilde{z}}\psi$ and Hamiltonian $H = \left(2\tilde{r}\right)^{-1}\sqrt{p_{\tilde{r}}^2 + p_{\tilde{z}}^2}$ . Then (11) takes the Hamilton-Jacobi form $p_\tau + H = 0$ for $p_\tau \equiv \partial_\tau\psi$. The Hamiltonian equations for the characteristic line elements are then

$$\frac{d\tilde{r}}{d\tau} = \frac{\partial H}{\partial p_{\tilde{r}}} = \frac{p_{\tilde{r}}}{2\tilde{r}\sqrt{p_{\tilde{r}}{}^2 + p_{\tilde{z}}{}^2}} = \frac{p_{\tilde{r}}}{4\tilde{r}^2 H} \tag{12}$$

$$\frac{d\tilde{z}}{d\tau} = \frac{\partial H}{\partial p_{\tilde{z}}} = \frac{p_{\tilde{z}}}{2\tilde{r}\sqrt{p_{\tilde{r}}{}^2 + p_{\tilde{z}}{}^2}} = \frac{p_{\tilde{z}}}{4\tilde{r}^2 H} \tag{13}$$

$$\frac{dp_{\tilde{r}}}{d\tau} = -\frac{\partial H}{\partial \tilde{r}} = \sqrt{p_{\tilde{r}}{}^2 + p_{\tilde{z}}{}^2}\,\frac{1}{2\tilde{r}^2} = \frac{H}{\tilde{r}} \tag{14}$$

$$\frac{dp_{\tilde{z}}}{d\tau} = -\frac{\partial H}{\partial \tilde{z}} = 0 \tag{15}$$

This system of equations has two invariants. Equation (15) shows $p_{\tilde{z}}$ to be one; H can be shown to be the other [1]. Combine the two invariants to define a new invariant N:

$$N \equiv \frac{p_{\tilde{z}}}{2H} \tag{16}$$

The definition of H then gives[1]

$$p_{\tilde{z}} = s\frac{Np_{\tilde{r}}}{\sqrt{\left(\tilde{r}^2 - N^2\right)}}, s = \pm 1; \tag{17}$$

Equations (12) and (13) then give

$$\frac{d\tilde{z}}{d\tau} = \frac{p_{\tilde{z}}}{4\tilde{r}^2 H} = \frac{N}{2\tilde{r}^2} \tag{18}$$

$$\frac{d\tilde{r}}{d\tau} = \frac{p_{\tilde{r}}}{4\tilde{r}^2 H} = \frac{s}{2\tilde{r}^2}\sqrt{\left(\tilde{r}^2 - N^2\right)} \tag{19}$$

Combining (18) and (19), the cosine of the angle $\phi$ made by the local normal to the GV surface with the z axis is given by

$$\cos\phi = \frac{d\tilde{z}/d\tau}{\sqrt{\left(d\tilde{z}/d\tau\right)^2 + \left(d\tilde{r}/d\tau\right)^2}} = \tilde{r}^{-1}N \tag{20}$$

Equation (20) provides a geometrical interpretation to the quantity N. It also shows that N is always less than $\tilde{r}$ and N is positive when $\phi$ is either in the first or the 4th quadrant. This condition can fail for a cavity in the anode profile and when the solution travels in the negative $\tilde{z}$ direction. The equation of the CLE is obtained from

$$\frac{d\tilde{z}}{d\tilde{r}} = \frac{d\tilde{z}}{d\tau}\Big/\frac{d\tilde{r}}{d\tau} = \frac{sN}{\sqrt{\tilde{r}^2 - N^2}} \tag{21}$$

For $N \neq 0$, integration of (21) gives the relation describing the family $\mathbb{C}$ of curves of constant N and H, which are perpendicular to the GV surface[1,4]:

$$\frac{\tilde{z}}{N} + s\mathrm{ArcCosh}\left(\frac{\tilde{r}}{|N|}\right) = C_1 = \mathrm{constant} \tag{22}$$

Integration of (19) gives the location of the integral surface on the family $\mathbb{C}$ of curves at time $\tau$

$$\frac{\tilde{r}}{N}\sqrt{\frac{\tilde{r}^2}{N^2} - 1} + \mathrm{ArcCosh}\left(\frac{\tilde{r}}{|N|}\right) + \frac{s\tau}{N^2} = C_2 = \mathrm{constant} \tag{23}$$

For the case of N=0, (19) shows

$$\tilde{r}^2 = s\tau + C_3, \tag{24}$$

The constants of integration $C_1, C_2, C_3$ are to be determined by the specifications of the problem as discussed below. The sign s is determined from the physical context. In addition, the solution must satisfy supplementary conditions mentioned above. The nature of the characteristic curves and GV surface for the case of Mather type plasma focus has been discussed in detail in Ref. 5. Note that equations (22) and (23) do not admit any scaling other than the normalization of the coordinates by the scale length 'a', which, however, is completely arbitrary.

The first supplementary condition mentioned above implies that the characteristic should be perpendicular to the given insulator profile at its intersection. From (21) ,

$$\frac{d\tilde{R}_I(\tilde{Z})}{d\tilde{Z}} = -\frac{d\tilde{z}}{d\tilde{r}} = -s\frac{N_I(\tilde{Z})}{\sqrt{\tilde{R}_I^{\,2}(\tilde{Z}) - N_I^{\,2}(\tilde{Z})}} \tag{25}$$

Therefore, in the insulator (or plasma formation) region

$$N_I(\tilde{Z}) = \frac{\tilde{R}_I(\tilde{Z})\left|d\tilde{R}_I(\tilde{Z})/d\tilde{Z}\right|}{\sqrt{\left(d\tilde{R}_I(\tilde{Z})/d\tilde{Z}\right)^2 + 1}} \tag{26}$$

Similarly, the third supplementary condition implies that the characteristic should be tangent to the anode profile at its intersection. Using (21)

$$\frac{d\tilde{r}}{d\tilde{z}} = s\frac{\sqrt{\tilde{r}^2 - N_A^{\,2}(\tilde{Z})}}{N_A(\tilde{Z})} = \frac{d\tilde{R}_A(\tilde{Z})}{d\tilde{Z}} \tag{27}$$

This gives

$$N_A(\tilde{Z}) = S\frac{\tilde{R}_A(\tilde{Z})}{\sqrt{1 + \left(d\tilde{R}_A/d\tilde{Z}\right)^2}}; S = \pm 1 \tag{28}$$

The second supplementary condition implies that the point of intersection between the GV surface and anode profile in the $(\tilde{r},\tilde{z})$ plane must move along the anode profile as a function of $\tau$ as described by (18), where $\tilde{r} = \tilde{R}_A(\tilde{Z})$ and $\tilde{z} = \tilde{Z}$ :

$$\frac{d\tilde{Z}}{d\tau} = \frac{N_A(\tilde{Z})}{2\tilde{R}_A^{\,2}(\tilde{Z})} \tag{29}$$

From (29), it is clear that the plus sign must be chosen in (28) when the solution is expected to begin at $\tilde{Z}=0$ and move towards higher $\tilde{Z}$ as in the case of the external branch of anode profile. There may be situations such as a hollow anode or a cavity in the anode face, (represented by the internal branch in anode profile function (1)) when the solution is expected to move within the cavity towards decreasing values of $\tilde{Z}$. In such case, the negative sign would apply in (28). One could then set $S=\mathrm{Sign}\left[d\tilde{Z}/d\tau\right]$.

Integration of (29) using (28) gives the value of $\tau\left(\tilde{Z}\right)$ when the GV surface reaches the point $P_A\left(\tilde{R}_A\left(\tilde{Z}\right),\tilde{Z}\right)$ on the external branch of anode profile

$$\tau\left(\tilde{Z}\right)-\tau_{i-1}=2S\int_{\tilde{Z}_A^{(i-1)}}^{\tilde{Z}}\sqrt{1+\left(d\tilde{R}_A^{(i)}/d\tilde{Z}\right)^2}\,\tilde{R}_A^{(i)}\left(\tilde{Z}\right)d\tilde{Z};\quad \tilde{Z}_A^{(i-1)}\le\tilde{Z}\left(\tau\right)\le\tilde{Z}_A^{(i)};\tau_0=0 \tag{30}$$

The dimensionless time $\tau_i$ at which the GV surface reaches the vertex $\tilde{Z}_A^{(i)}$ is given by the recursion equations

$$\tau_i-\tau_{i-1}=2S\int_{\tilde{Z}_A^{(i-1)}}^{\tilde{Z}_A^{(i)}}\sqrt{1+\left(d\tilde{R}_A^{(i)}/d\tilde{Z}\right)^2}\,\tilde{R}_A^{(i)}\left(\tilde{Z}\right)d\tilde{Z} \tag{31}$$

The function $\tau\left(\tilde{Z}\right)$ may be numerically inverted to obtain the coordinates of the intersection of GV surface with the external anode surface for a given $\tau$. For the RGV model case applicable to Mather type plasma focus, (31) reproduces the expression for the rundown time as $\tau_R=2\left(\tilde{z}_A-\tilde{z}_I\right)$.

It is important to note that the function $\tau\left(\tilde{Z}\right)$ is a monotonically increasing function within each segment of external branch of anode profile. It implies that curves of family $\mathbb{C}$ originating on the external branch of the anode profile within the space bounded by the cathode are ordered in dimensionless time $\tau$– those that begin from a point $P_A\left(\tilde{R}_A\left(\tilde{Z}\right),\tilde{Z}\right)$ at $\tau\ge\tau\left(\tilde{Z}\right)$ are properly identified as *characteristic* curves within the meaning of the Method of Characteristics [21]. The constant $C_1$ for each characteristic curve is then given by

$$C_{1,A}\left(\tilde{Z},N\right)=\frac{\tilde{Z}}{N_A\left(\tilde{Z}\right)}+s\,\mathrm{ArcCosh}\left(\frac{\tilde{R}_A\left(\tilde{Z}\right)}{\left|N_A\left(\tilde{Z}\right)\right|}\right) \tag{32}$$

Varying $\tilde{r}$ in (22) from $\tilde{R}_A\left(\tilde{Z}\right)$ to $\tilde{r}_C$ with $N=N_A\left(\tilde{Z}\right)$ generates the characteristic curve $\left(\tilde{r}_{ch},\tilde{z}_{ch}\right)$ for point $P_A\left(\tilde{R}_A\left(\tilde{z}\right),\tilde{z}\right)$. At the vertices $\left(\tilde{R}_A\left(\tilde{Z}_A^{(i)}\right),\tilde{Z}_A^{(i)}\right)$, $d\tilde{R}_A/d\tilde{Z}$ may be discontinuous. Characteristics at these points are found by using

$$C_{1,A}\left(\tilde{Z}_A^{(i)}\right)=\frac{\tilde{Z}_A^{(i)}}{N}+s\,\mathrm{ArcCosh}\left(\frac{\tilde{R}_A\left(\tilde{Z}_A^{(i)}\right)}{|N|}\right) \tag{33}$$

by varying N continuously from

$$N_A^-\left(\tilde{Z}_A^{(i)}\right)=\tilde{R}_A\left(\tilde{Z}_A^{(i)}\right)\Big/\sqrt{1+\left(d\tilde{R}_A^{(i-1)}\left(\tilde{Z}\right)/\tilde{Z}\right)^2_{\tilde{Z}\to\tilde{Z}_A^{(i)}}} \tag{34}$$

to

$$N_A^+\left(\tilde{Z}_A^{(i)}\right)=\tilde{R}_A\left(\tilde{Z}_A^{(i)}\right)\Big/\sqrt{1+\left(d\tilde{R}_A^{(i)}\left(\tilde{Z}\right)/\tilde{Z}\right)^2_{\tilde{Z}\to\tilde{Z}_A^{(i)}}}\,. \tag{35}$$

Each characteristic is therefore labeled *both* by its starting point $P_A\left(\tilde{R}_A\left(\tilde{z}\right),\tilde{z}\right)$ and its N value as $Ch\left(\tilde{Z},N\right)$.

Similarly, every point $P_I\left(\tilde{R}_I\left(\tilde{z}\right),\tilde{z}\right)$ on the insulator generates a characteristic curve whose constant $C_1$ is given by

$$C_{1,In}\left(\tilde{Z}\right)=\pm\frac{\tilde{Z}\sqrt{\left(d\tilde{R}_I\left(\tilde{Z}\right)/d\tilde{Z}\right)^2+1}}{\tilde{R}_I\left(\tilde{Z}\right)\left(d\tilde{R}_I\left(\tilde{Z}\right)/d\tilde{Z}\right)}+s\,\mathrm{ArcCosh}\left(\frac{\sqrt{\left(d\tilde{R}_I\left(\tilde{Z}\right)/d\tilde{Z}\right)^2+1}}{\left|\left(d\tilde{R}_I\left(\tilde{Z}\right)/d\tilde{Z}\right)\right|}\right) \tag{36}$$

It is well known [21] that a single-valued, continuous and differentiable (classical) solution to a nonlinear first order partial differential equation exists if and only if the characteristic curves cover the entire domain and do not intersect. For arbitrary anode and insulator shapes, this circumstance is not guaranteed. Intersection of characteristic curves or their exclusion from certain regions of the solution domain leads to non-classical solutions that represent a system of shock waves and rarefaction waves where additional physical conditions such as the "entropy condition" or a "vanishing viscosity" need to be invoked to decide which is a "physical solution".

However, if the intersecting curves of family $\mathbb{C}$ are associated with different values of $\tau$, then those that correspond to earlier time are to be considered the "genuine" characteristics and those that correspond to later time are "forbidden to intrude in the region" occupied by the former. *This is because the characteristic curves are collection of characteristic line elements defined in the neighborhood of every point on the GV surface at a given time including the initial time*. This process prevents generation of intersecting *characteristics* even though intersecting curves obeying the algebraic equation (22) may be drawn over the anode profile. The solution region corresponding to each segment of external branch of anode profile should thus consist of three (or more) sub-regions: one which is based on characteristics drawn from the points on the external branch of anode profile in that segment, second which is based on characteristics drawn from the vertex at the beginning of the segment and the third that is based on characteristics drawn from points in the previous anode segment or insulator (with additional regions based on characteristics drawn from still earlier segments). This is illustrated in Section III. The characteristics associated with the internal branch of anode profile do not intersect with those associated with the external branch except at the transition zone between the two.

For the purpose of the present discussion, it is assumed that anode and insulator profiles have been chosen in such manner that the characteristics associated with same value of time do not intersect and that a classical solution to the GV equation exists in each sub-region of the solution space as discussed above.

The GV surface at dimensionless time $\tau$ is the locus of those points on the entire family $\mathbb{C}$ of characteristic curves, whose radial coordinates satisfy equation (23), and which continuously connect the anode with the cathode. The dimensionless time $\tau$ at which the intersection of the GV surface with the anode reaches the point $P_A\left(\tilde{R}_A\left(\tilde{Z}\right),\tilde{Z}\right)$ has already been determined above.

The constant $C_2$ in (23) for the characteristic $Ch\left(\tilde{Z},N\right)$ is then given by

$$C_{2,A}\left(\tilde{Z},N\right)=\frac{\tilde{R}_A\left(\tilde{Z}\right)}{N}\sqrt{\frac{\tilde{R}_A{}^2\left(\tilde{Z}\right)}{N^2}-1}+\operatorname{ArcCosh}\left(\frac{\tilde{R}_A\left(\tilde{Z}\right)}{\left|N\right|}\right)+\frac{s\tau\left(\tilde{Z}\right)}{N^2} \tag{37}$$

At any time $\tau>\tau\left(\tilde{Z}\right)$ , the intersection of the GV surface with the characteristic $Ch\left(\tilde{Z},N\right)$ can be calculated by solving equation (23) for $\tilde{r}\equiv\tilde{r}_{GV}\left(\tau,\tilde{Z},N\right)$, with $C_2$ given by (37) and substituting in (22) for obtaining $\tilde{z}\equiv\tilde{z}_{GV}\left(\tau,\tilde{Z},N\right)$. The set of all such points for a given $\tau$ is the GV surface.

GV have given the following method of calculating the GV surface. Rewrite (22), the equation for the characteristic $Ch\left(\tilde{Z},N\right)$, defining $\alpha/2=C_1-\tilde{z}/N$, $\beta/2\equiv\operatorname{ArcCosh}\left(\tilde{R}_A\left(\tilde{Z}\right)/N\right)$ as

$$\begin{aligned}&\tilde{r}_{GV}\left(\alpha,N\right)=N\operatorname{Cosh}\left(\alpha/2\right);\\&\tilde{z}_{GV}\left(\alpha,N,s\right)=NC_1\left(\tilde{Z},N\right)-Ns\,\alpha/2\end{aligned} \tag{38}$$

Then (23) can be written in three different ways as

$$\begin{aligned}F\left(\alpha\right)&\equiv\operatorname{Sinh}\left(\alpha\right)+\alpha=2\left(C_2\left(\tilde{Z},N,s\right)-\frac{s\tau}{N^2}\right) &&\cdots(a)\\&=2\left(\frac{\tilde{R}_A\left(\tilde{Z}\right)}{N\left(\tilde{Z}\right)}\sqrt{\frac{\tilde{R}_A\left(\tilde{Z}\right)^2}{N^2\left(\tilde{Z}\right)}-1}+\operatorname{ArcCosh}\left(\frac{\tilde{R}_A\left(\tilde{Z}\right)}{N\left(\tilde{Z}\right)}\right)-s\frac{\left(\tau-\tau\left(\tilde{Z}\right)\right)}{N^2\left(\tilde{Z}\right)}\right) &&\cdots(b)\\&=\operatorname{Sinh}\left(\beta\right)+\beta-\operatorname{Cosh}^2\left(\beta/2\right)\frac{2s\left(\tau-\tau\left(\tilde{Z}\right)\right)}{\tilde{R}_A\left(\tilde{Z}\right)^2} &&\cdots(c)\end{aligned} \tag{39}$$

The function $F\left(\alpha\right)$ is an odd function of its argument. GV provide [1] the following inversion formula for $\left|F\left(\alpha\right)\right|$

$$\begin{aligned}&\left|\alpha\right|=\frac{\left|F\right|}{2}\left(1-\frac{\left|F\right|^2}{48}\right) \qquad \text{for} \quad \left|F\right|<1\\&\left|\alpha\right|=\operatorname{Log}\left[2\left|F\right|+\frac{1}{2\left|F\right|}-2\left(1-\frac{1}{\left|F\right|}\right)\operatorname{Log}\left(2\left|F\right|\right)\right] \quad \text{for} \quad \left|F\right|>>1\end{aligned} \tag{40}$$

GV quote an error of less than 0.1% in the second formula for $|F| > 5$. For the entire set of characteristics $Ch(\tilde{Z}, N)$, $\alpha$ can be determined as a function of $\tau$ from equation (39)(a) from which the GV surface $(\tilde{r}_{GV}, \tilde{z}_{GV})$ is computed from (38).

For the special case of constant anode radius (e.g. rundown phase in the Mather type plasma focus), (28) gives $N_A(\tilde{Z}) = \tilde{R}_A$ . Then $C_{1,A}(\tilde{Z}, N) = \tilde{Z}/\tilde{R}_A$, $C_{2,A}(\tilde{Z}, N) = s\tau(\tilde{Z})/\tilde{R}_A{}^2$. From (30), $\tau(\tilde{Z}) = 2\tilde{R}_A(\tilde{Z} - \tilde{Z}_A^{(i)}) + \tau_i$. This can be used to eliminate $\tilde{Z}$ in (22) and (23) giving

$$\tilde{z} = \tilde{Z}_A^{(i)} + \frac{1}{2\tilde{R}_A}(\tau - \tau_i) + \frac{1}{2}\tilde{R}_A \mathrm{ArcCosh}\left(\frac{\tilde{r}}{\tilde{R}_A}\right) - \frac{1}{2}\frac{\tilde{r}}{\tilde{R}_A}\sqrt{\tilde{r}^2 - \tilde{R}_A{}^2} \tag{41}$$

Varying $\tilde{r}$ from $\tilde{R}_A$ to $\tilde{r}_C$ generates the GV surface in the segment with constant anode radius *if characteristics from the previous anode profile segment do not intrude into the rundown space*.

The solution of the generalized plasma focus problem (i.e. GV equation along with the supplementary conditions) requires appreciation of the existence of an inherent bifurcation phenomenon. The solution to the problem consists of the coordinates $(\tilde{r}_{GV}(\alpha, N), \tilde{z}_{GV}(\alpha, N, s))$ defined by (38), where $\alpha$ varies with N at a given time $\tau$ according to (39). After a monotonic variation, $\alpha$ may cross zero and change its sign. Beyond the condition defined by $\alpha = 0$ the solution breaks up into two branches which correspond to $\tilde{r} = N\mathrm{Cosh}(|\alpha|/2)$, $\tilde{z} = NC_1(\tilde{Z}, N) \pm Ns|\alpha|/2$. From (39)

$$\alpha = 0 \Rightarrow \frac{\tilde{R}_A(\tilde{Z})}{N(\tilde{Z})}\sqrt{\frac{\tilde{R}_A(\tilde{Z})^2}{N^2(\tilde{Z})} - 1} + \mathrm{ArcCosh}\left(\frac{\tilde{R}_A(\tilde{Z})}{N(\tilde{Z})}\right) = s\frac{(\tau - \tau(\tilde{Z}))}{N^2(\tilde{Z})} \tag{42}$$

At the vertex $\tilde{Z} = \tilde{Z}_A^{(i)}$, as N becomes decoupled from $\tilde{Z}$ and takes a continuous range of values between $N_A^-(\tilde{Z}_A^{(i)})$ and $N_A^+(\tilde{Z}_A^{(i)})$, (42) becomes an equation whose LHS is positive definite. The RHS is also positive for s=+1 and $(\tau - \tau_i) > 0$ so that a real solution $N_{br}(\tau - \tau_i)$ to (42) might exist. The coordinates $(\tilde{r}_{br}, \tilde{z}_{br})$ of the branch points are then given by

$$\tilde{r}_{br} = N_{br}\left(\tau-\tau_i\right);\ \tilde{z}_{br} = \tilde{Z}_A^{(i)} + N_{br}\left(\tau-\tau_i\right)\mathrm{ArcCosh}\left(\frac{\tilde{R}_A\left(\tilde{Z}_A^{(i)}\right)}{N_{br}\left(\tau-\tau_i\right)}\right) \tag{43}$$

Note that at the branch points, (21) shows that

$$\left.\frac{d\tilde{r}}{d\tilde{z}}\right|_{\tilde{r}_{br}} = s\frac{\sqrt{\tilde{r}_{br}{}^2 - N^2}}{N} = 0 \tag{44}$$

The branch point radius $\tilde{r}_{br}$ is thus the minimum possible radius of the GV surface at $\tau$ for the concerned vertex or segment.

For the segment $\tilde{R}_{A,ext}^{(i)}\left(\tilde{Z}\right)$, $\tilde{Z}_A^{(i-1)} \le \tilde{Z} \le \tilde{Z}_A^{(i)}$, the GV surfaces are parametrically dependent on $\tilde{Z}$. Extrema of $\left(\tilde{r}_{GV}\left(\tilde{Z},\tau\right),\tilde{z}_{GV}\left(\tilde{Z},\tau\right)\right)$ may exist that define additional branch points dependent on the anode profile.

Another notable phenomenon that needs to be appreciated is that the family $\mathbb{V}$ of characteristics emitted from a vertex $\left(\tilde{r}_v,\tilde{z}_v\right)$ given by $\tilde{r} = \tilde{r}\left(\alpha,N\right);\tilde{z} = \tilde{z}\left(\alpha,N,s,\tilde{r}_v,\tilde{z}_v\right)$ has an envelope $\Sigma_v$ (a curve that is tangent to each member of the family $\mathbb{V}$ ) which satisfies the equation[22]

$$\partial_\alpha\tilde{r}\left(\alpha,N\right)\partial_N\tilde{z}\left(\alpha,N,s,\tilde{r}_v,\tilde{z}_v\right) - \partial_N\tilde{r}\left(\alpha,N\right)\partial_\alpha\tilde{z}\left(\alpha,N,s,\tilde{r}_v,\tilde{z}_v\right) = 0 \tag{45}$$

This gives $\alpha_{sol}\left(N/\tilde{r}_v\right)$ as a function of $\left(N/\tilde{r}_v\right) = \mathrm{Sech}\left(\beta/2\right)$ as a solution of the following equation[23]

$$\mathrm{Coth}\left(\alpha/2\right) - \frac{1}{2}\alpha = \mathrm{Coth}\left(\beta/2\right) - \frac{1}{2}\beta \tag{46}$$

The envelope is then given by the parametric curve

$$\Sigma_v = \left(\tilde{r}\left(\alpha_{sol}\left(N/\tilde{r}_v\right),N\right),\tilde{z}\left(\alpha_{sol}\left(N/\tilde{r}_v\right),N,s,\tilde{r}_v,\tilde{z}_v\right)\right) \tag{47}$$

which explicitly demonstrates that the envelope satisfies equation (22) that defines characteristics of the GV equation and is therefore itself a characteristic curve. The envelope and the characteristics are shown in Fig 1.

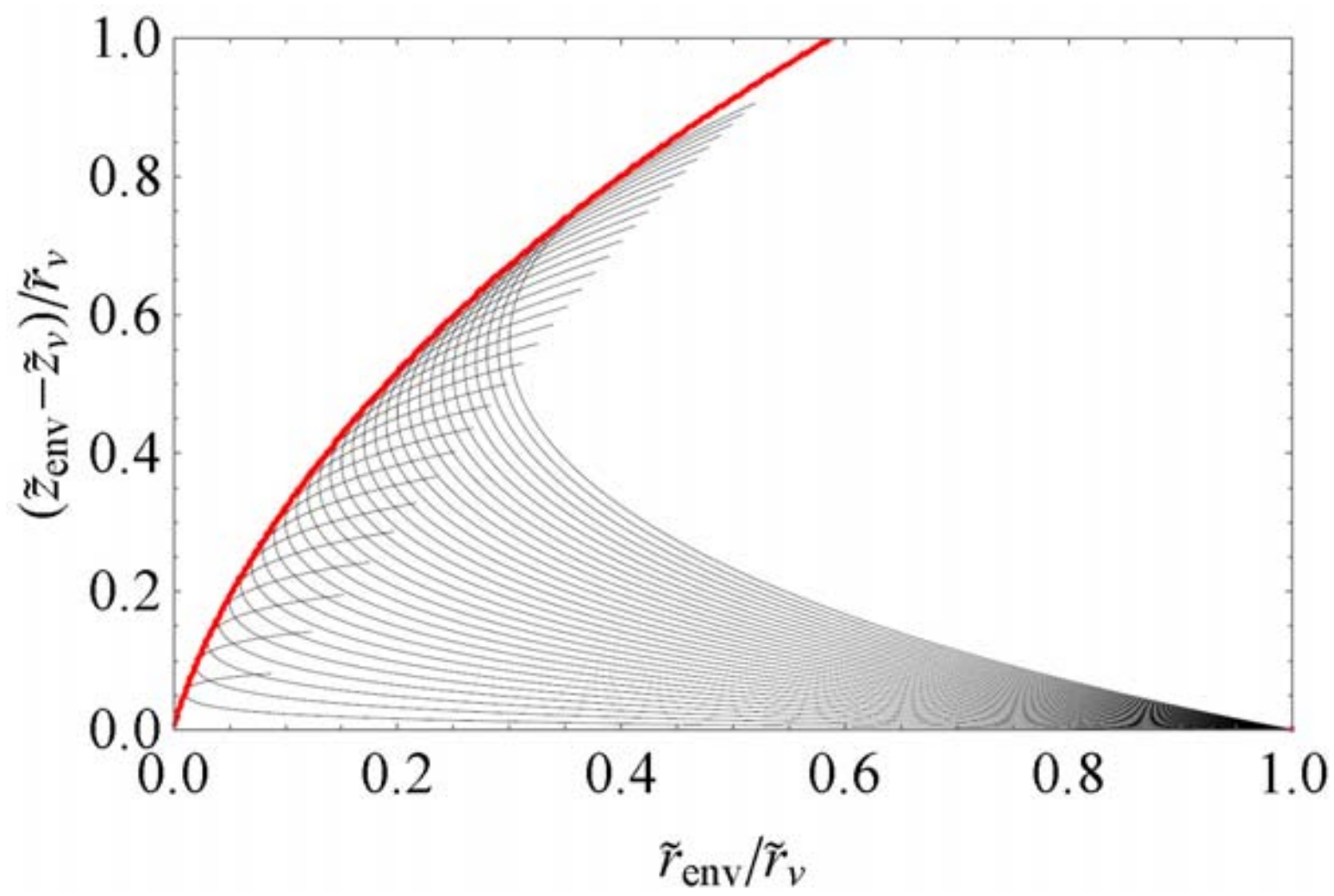

Fig. 1: Family $\mathbb{V}$ of characteristics emitted from vertex $(\tilde{r}_v, \tilde{z}_v)$ with s=1 shown in black, parametrized by N. For each curve, N is constant while $\alpha$ varies along the curve. The envelope $\Sigma_v$ is shown as a thick red line. Note that the envelope represents the boundary of the zone of influence of the family $\mathbb{V}$ of characteristics starting from vertex $(\tilde{r}_v, \tilde{z}_v)$.

Since the envelope is obviously the locus of smallest-radius points on the family of characteristics, it is also the locus of all branch points described by(43).

It is clear that the vertex $(\tilde{r}_v, \tilde{z}_v)$ influences $\Sigma_v$ in only two ways: (1) by providing a scaling of radial and axial coordinates of the envelope with $\tilde{r}_v$ and (2) by redefining origin for the axial coordinate of the envelope with $\tilde{z}_v$. The shape of the function $(\tilde{z}_{\text{env}} - \tilde{z}_v)/\tilde{r}_v$ versus $\tilde{r}_{\text{env}}/\tilde{r}_v$ is in fact universal. Fig 2 compares a power-law fit with this shape

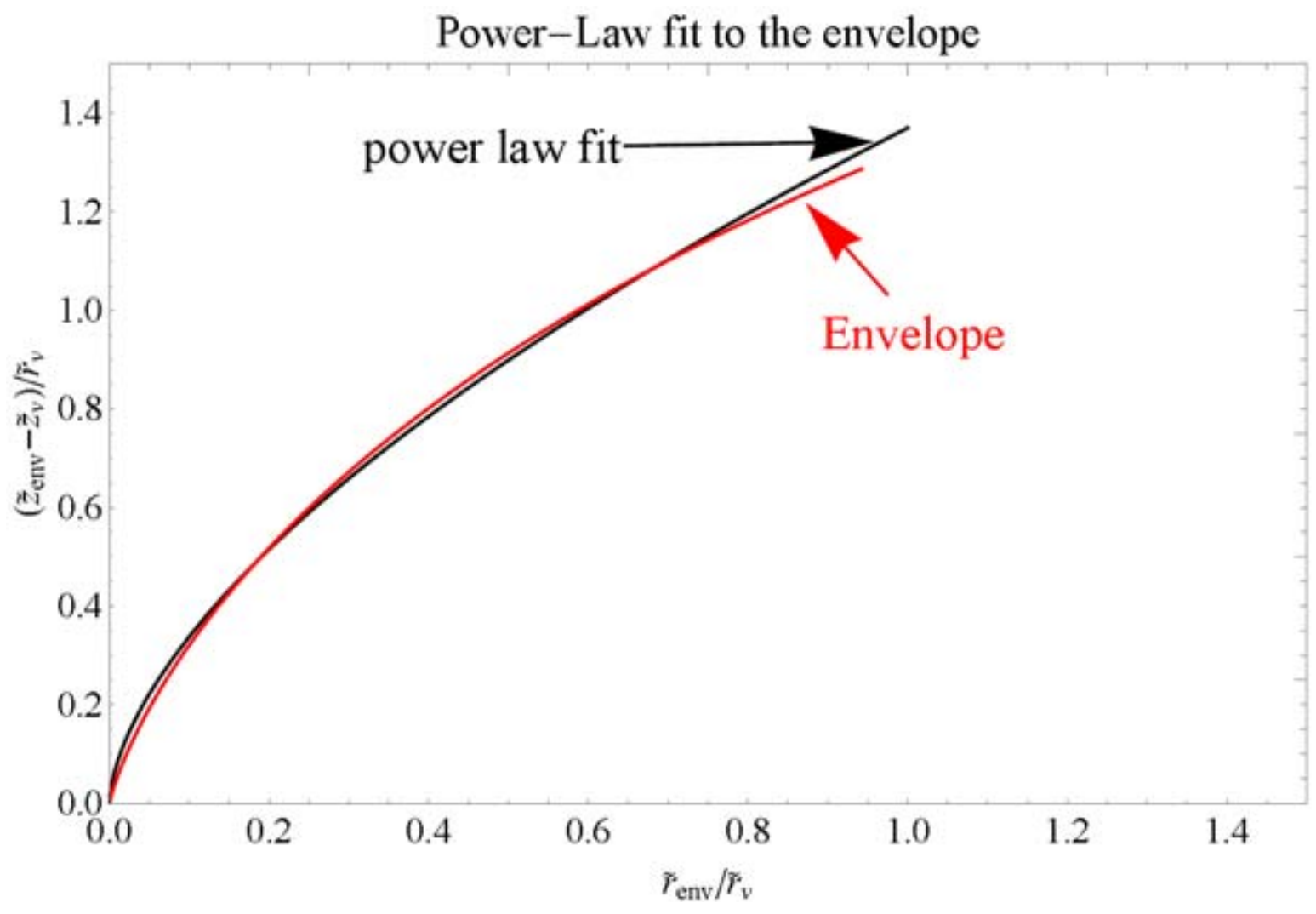

Fig 2: Power law fit to the envelope $\Sigma_v : \ (\tilde{z}_{env} - \tilde{z}_v)/\tilde{r}_v = a_1 \times (\tilde{r}_{env}/\tilde{r}_v)^{a_2} \ \ a_1 = 1.36996; a_2 = 0.607746$. The fit constants change if only a part of the envelope profile is chosen to obtain a better quality of fit.

In the case of the generalized plasma focus problem, the characteristics in the segment $\tilde{R}^{(i)}_{A,ext}(\tilde{Z}), \quad \tilde{Z}^{(i-1)}_{A} \le \tilde{Z} \le \tilde{Z}^{(i)}_{A}$ are given by

$$\begin{aligned} \tilde{r}(\alpha,\tilde{Z}) &= N_A(\tilde{Z})\,\mathrm{Cosh}(\alpha/2) \\ \tilde{z}(\alpha,\tilde{Z},s) &= \tilde{Z} + sN_A(\tilde{Z})\,\mathrm{ArcCosh}\left(\frac{\tilde{R}_A(\tilde{Z})}{N_A(\tilde{Z})}\right) - N_A(\tilde{Z})\,s\,\alpha/2 \end{aligned} \tag{48}$$

The envelope of characteristics is then given by[22]

$$\partial_\alpha \tilde{r}(\alpha,\tilde{Z})\,\partial_{\tilde{Z}}\tilde{z}(\alpha,\tilde{Z},s) - \partial_{\tilde{Z}}\tilde{r}(\alpha,\tilde{Z})\,\partial_\alpha \tilde{z}(\alpha,\tilde{Z},s) = 0 \tag{49}$$

whose numerical solution for a given anode profile gives $\alpha_{sol}(\tilde{Z})$ from which the envelope is obtained as

$$\Sigma_s = \left(\tilde{r}\left(\alpha_{sol}(\tilde{Z}),\tilde{Z}\right), \tilde{z}\left(\alpha_{sol}(\tilde{Z}),\tilde{Z},s\right)\right) \tag{50}$$

Since the characteristics are tangent to the anode profile, in many cases, the envelope simply turns out to be the anode profile. However, it is not known whether any characteristics other than the anode profile may exist for certain profile functions.

The properties of the GV surface in the limit $N \to 0$ merit a special discussion. One important example of such case is the neighborhood of the axis. From (38), it is clear that the GV surface can reach the axis at $\tilde{r} = 0$ only when $N \to 0$, since the hyperbolic cosine function cannot become zero. However, the right hand side of (39)(b) diverges as $N \to 0$: it has the following series expansion in powers of N

$$\frac{2\left(\left(\tilde{R}_A\left(\tilde{Z}\right)\right)^2 - s\left(\tau - \tau\left(\tilde{Z}\right)\right)\right)}{N^2} - 1 + \log\left[\frac{4\left(\tilde{R}_A\left(\tilde{Z}\right)\right)^2}{N^2}\right] - \frac{3N^2}{4\left(\tilde{R}_A\left(\tilde{Z}\right)\right)^2} - \frac{5N^4}{16\left(\tilde{R}_A\left(\tilde{Z}\right)\right)^4} + O[N]^6 \quad (51)$$

showing that there is an essential singularity that vanishes when s=+1 and

$$\tau \to \tau\left(\tilde{Z}\right) + \left(\tilde{R}_A\left(\tilde{Z}\right)\right)^2 \equiv \tau_p \quad (52)$$

Additionally, there is a logarithmic singularity.

Since the right hand side of (39) diverges as $N \to 0$, the second expression of (40) applies and can be approximated for large $\left|F\left(\alpha\right)\right|$ as $\left|\alpha\right| \approx \mathrm{Log}\left[\left|2F\right|\right]$, $\mathrm{Cosh}\left(\alpha/2\right) \approx \frac{1}{2}\exp\left(\alpha/2\right) \approx \sqrt{\left|F\right|/2}$ . Then (38) gives

$$\tilde{r}^2 = N^2 \mathrm{Cosh}^2\left(\alpha/2\right)$$
$$\approx \left|\left(\tilde{R}_A\left(\tilde{Z}\right)\right)^2 - s\left(\tau - \tau\left(\tilde{Z}\right)\right) - N^2 + N^2\log\left[\frac{4\left(\tilde{R}_A\left(\tilde{Z}\right)\right)^2}{N^2}\right] - N^2\left\{\frac{3N^2}{4\left(\tilde{R}_A\left(\tilde{Z}\right)\right)^2} + \frac{54N^4}{16\left(\tilde{R}_A\left(\tilde{Z}\right)\right)^4} + O[N]^7\right\}\right| \quad (53)$$

Two cases need to be discussed separately: (I) Type I anode profile where the normalized radius of the last "flat-top" segment of the external branch of anode profile (before beginning of the cavity if any), $\tilde{r}_{last} = \tilde{R}_A^{(last)}\left(\tilde{Z}_A^{(last)}\right)$ is assumed to be non-zero (it may be arbitrarily small), (Note that according to the discussion of (28), N has negative values in the cavity.) (II) Type II anode profile where $R_A^{(last)}(\tilde{Z}) \to 0$ as $\tilde{Z} \to \tilde{z}_A$ where $\tilde{Z}_A^{(last)}$ is the last vertex of the anode profile before $\tilde{z}_A$. The difference in the two cases lies in the fact that $\tilde{Z}$ ceases to be a good variable to describe points on the flat top anode in equations (32) or (37) in the first case but not the second case. The characteristics that enable the GV surface to reach the axis are emitted from the last vertex in the first case but are emitted from each point on the anode profile right up to the axis. So for the first case, functions $\tilde{R}_A(\tilde{Z})$ and $\tau(\tilde{Z})$ are replaced with constants $\tilde{r}_{last}$ and $\tau_{last}$ in (53). From (28), it is seen that the ratio $\tilde{R}_A(\tilde{Z})/N$ in (53) has a finite limit in the second case as $N \to 0$.

In both cases, it is clear that the last 3 terms of (53) vanish in the limit of $N \to 0$ but the variation of $\tilde{r}^2$ with sufficiently small values of N is different for the two cases. The result for $N \to 0$ is

$$\tilde{r}^2 = \left|\tilde{r}_{last}{}^2 - s(\tau - \tau_{last})\right| \qquad \text{or } \tilde{r}^2 = \left|\left(\tilde{R}_A(\tilde{Z})\right)^2 - s\left(\tau - \tau(\tilde{Z})\right)\right|, \quad \tilde{Z}_A^{(i_m-1)} \le \tilde{Z} \le \tilde{Z}_A^{(i_m)} \qquad (54)$$

Since $\tau > \tau_{last}\left(\text{or } \tau\left(\tilde{Z}_A^{(i_m-1)}\right)\right)$, the axis is approached only when s=+1. Also, from (38) and (33), $\tilde{z} \to \tilde{z}_A$ as $N \to 0$. This shows that *in the plane containing the top of the anode, the GV surface resembles contracting or expanding circles centered at the axis*.

The fact that identical non-zero values of the radial coordinate occur at two values of $\tau$ implies that the GV surface gets reflected from the axis at $\tau = \tau_{last} + \tilde{r}_{last}{}^2 \equiv \tau_p$ for the first case or $\tau = \tau(\tilde{z}_A) \equiv \tau_p$ for the second case. *This follows purely from the mathematical properties of the GV equation and has nothing to do with the physics of shock wave reflection from a symmetry axis that involves conservation laws for mass, momentum and energy*.

This reflection is in fact an instance of the bifurcation phenomenon described earlier, with the branching point at the axis in the plane containing the vertex $\tilde{Z}_A^{(last)}$. All subsequent branch points lie on the envelope of characteristics emitted from $\tilde{Z}_A^{(last)}$ for a Type I anode profile. The corresponding time $\tau_{br}$ can be obtained from (39) using the solution $\alpha_{sol}(N/\tilde{r}_{last})$ of (46):

$$\left(\tau_{br} - \tau_{last} - \tilde{r}_{last}{}^{2}\right)$$
$$= \tilde{r}_{last}{}^{2}\left\{\sqrt{1-\frac{N^2}{\tilde{r}_{last}{}^{2}}} - 1 + \frac{N^2}{\tilde{r}_{last}{}^{2}}\text{ArcCosh}\left(\frac{\tilde{r}_{last}}{N}\right) - \frac{N^2}{\tilde{r}_{last}{}^{2}}\frac{1}{2}\left(\text{Sinh}\left(\alpha_{sol}\left(N/\tilde{r}_{last}\right)\right) + \alpha_{sol}\left(N/\tilde{r}_{last}\right)\right)\right\} \quad (55)$$

The function in braces has a good regression fit to a 3rd degree polynomial in $\left(N/\tilde{r}_{last}\right)$ over the range $\left[0, 0.5\right]$ with coefficients $\{0, 0.454231, -0.879163, 8.11138\}$.

The lower branch spanning $0 \le N \le N_0\left(\tau_{br} - \tau_p\right)$, where $N = N_0\left(\tau_{br} - \tau_p\right)$ is the solution of (55), appears as a Radially Expanding Front (REF), that has the radius given by (54) in the plane containing the apex of the anode, which must further meet the anode. The upper branch spanning $N_0\left(\tau - \tau_p\right) \le N \le \tilde{r}_{last}$ forms part of the GV surface that connects the REF to the cathode.

For a Type II anode profile, branching of the GV surface occurs for $\tau > \tau_p$ but the locus of branch points is not given by (55). Since the details of the anode profile in the last segment play a role in determining the locus of branch points, one can simply determine it empirically for each case rather than analytically for a general profile, as done in the illustration in Section IV.

The space between the locus of branch points and the axis is not reached by any characteristic either from the vertex $\tilde{Z}_A^{(last)}$ for Type I anode profile or the last segment of the Type II anode profile. Since the envelope from the last vertex is shown to be a characteristic within the meaning of (22), which is tangent to the axis and additively dependent on the axial coordinate of the vertex, it can be used as a template for constructing a family $\mathbb{S}$ of characteristics by translating it along the axis with the free parameter $\tilde{z}_0$:

$$\tilde{\tilde{z}}_S = \tilde{\tilde{z}}_0 + a_1 \tilde{\tilde{r}}_S{}^{a_2} \quad (56)$$

The double overtilde in (56) denotes coordinates normalized to $\tilde{r}_{last}$ .

The GV surface in the space between the locus of branch points and the axis then consists of a family $\mathbb{O}$ of orthogonal curves of the family $\mathbb{S}$ given by ,

$$\tilde{\tilde{z}}_{GV} = a_3 - \frac{1}{a_1 a_2 \left(2 - a_2\right)} \tilde{\tilde{r}}_{GV}{}^{2-a_2} \qquad (57)$$

The constant $a_3$ is determined by matching (57) at the branch point given by (43) or otherwise determined for a given anode profile. The family $\mathbb{O}$ describes an Axially Expanding Front (AEF).

For times $\tau > \tau_p$, the GV surface therefore consists of *three* branches that intersect the locus of branch points at $\left(\tilde{r}_{br}, \tilde{z}_{br}\right)$: a Radially Expanding Front (REF), whose intersection with the anode has radius given by (54), which is orthogonal to the family $\mathbb{C}$; the Axially Expanding Front (AEF) that is orthogonal to the family $\mathbb{S}$ and a third branch that is orthogonal to $\mathbb{C}$ but which connects the point of intersection to the cathode (referred as CN),. This situation is illustrated in Fig 3 of Ref. 5 for a Mather type plasma focus.

For a Type II anode profile, the REF would be moving towards lower values of $\tilde{Z}$ so that $S = -1$ in (28) and (30). One can define a "ghost segment" on the external branch of anode profile for computational purposes to deal with the GV surface for $\tau > \tau_p$ with the sign of $N_A\left(\tilde{Z}\right)$ reversed. It should be noted that physical considerations of supplementary condition 3 do not apply to the reflected GV surface and there is no necessity that the reflected GV surface should intersect with the anode at right angles.

The AEF and REF define an enclosed volume that expands with dimensionless time $\tau > \tau_p$. Fig 3 illustrates this for the Mather geometry. Incidentally, $\left(\tilde{r}_{br}, \tilde{z}_{br}\right)$ also represent the radius and height of the largest cylinder that can be inscribed in this closed volume.

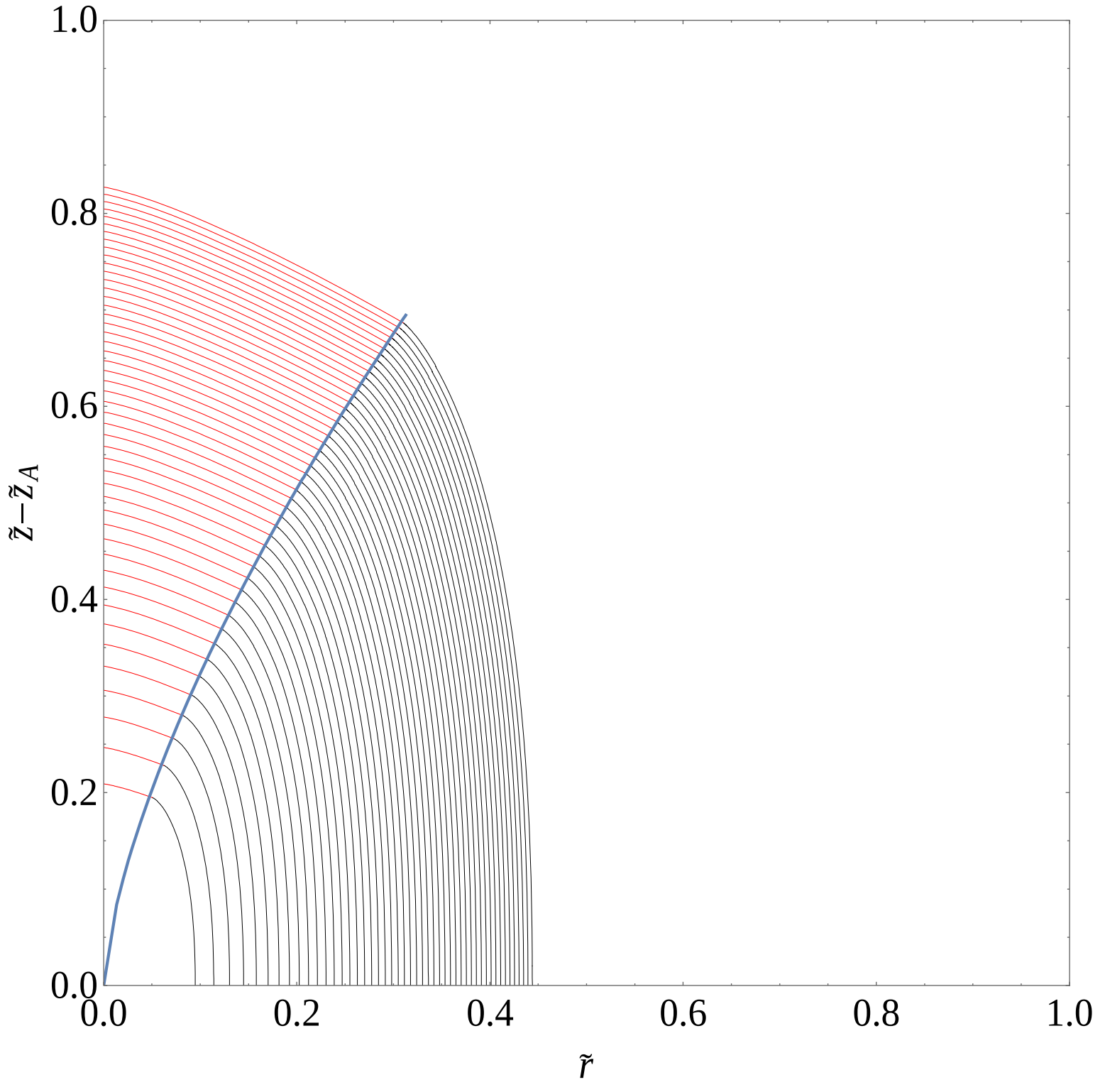

Fig 3.: The Axially Expanding Front (AEF), shown in red, and the Radially Expanding Front (REF), shown in black, define a closed volume that steadily expands for $\tau > \tau_p$. The figure is for the Mather geometry case. For any other case, the coordinates are scaled to the radius of the last vertex.

The point of intersection of the AEF and REF, $\left(\tilde{r}_{br}, \tilde{z}_{br}\right)$ , moves along the line shown in blue according to the parametric relation

$$\tilde{r}_{br} = 0.8143\left(\tau-\tau_p\right)^{0.59451}; \tilde{z}_{br} = 1.320\left(\tau-\tau_p\right)^{0.39898} \tag{58}$$

. Along the axis, the position of the APF is given by the regression fit

$$\tilde{z}_{APF} = 1.70246\left(\tau-\tau_p\right)^{0.443703} \tag{59}$$

The points on the AEF in Fig 2, are well-represented by the following regression fits

$$\tilde{r}_{APF} = \left(0.82227 - 0.00814\,j\right)\left(\tau - \tau_p\right)^{0.59392}$$
$$\tilde{z}_{APF} = \left(1.31164 + 0.00507\,j - 0.00001117\,j^2\right)\left(\tau - \tau_p\right)^{0.3982 + 0.0006571\,j - 0.000001994\,j^2} \qquad (60)$$

with j varying from 1 on the blue line separating the AEF and REF to 101 on the axis. No simple regression fit is available for the REF at this time.

The interesting point about this phenomenon is that the expanding enclosed volume has *no physics* associated with it other than what is built into the GV equation. It could be used to model the expansion of the DPF plasma after the pinch phase without invoking pressure balance and energy balance considerations, just to verify whether the experimentally observed plasma expansion is really sensitive to some physical assumptions or not.

It can be readily verified that the case of the classical Mather type plasma focus discussed earlier[1,4,5] is a special case of the above method, with 'a' chosen as the anode radius.

## III. Illustration with a non-standard geometry

This section offers a tutorial illustration of the procedure outlined above. For this purpose, the following profile is used

$$\begin{aligned}\tilde{R}_I\left(\tilde{Z}\right) &= \tilde{R}_I^{(0)} + \left(\tilde{R}_I^{(1)} - \tilde{R}_I^{(0)}\right)\tilde{Z}\Big/\tilde{Z}_I^{(1)} && 0 \le \tilde{Z} \le \tilde{Z}_I^{(1)} \Leftrightarrow I_1 \\ &= \tilde{R}_I^{(1)} + \left(\tilde{Z} - \tilde{Z}_I^{(1)}\right)\left(\tilde{R}_I^{(2)} - \tilde{R}_I^{(1)}\right)\Big/\left(\tilde{Z}_I^{(2)} - \tilde{Z}_I^{(1)}\right) && \tilde{Z}_I^{(1)} \le \tilde{Z} \le \tilde{Z}_I^{(2)} \Leftrightarrow I_2\end{aligned} \qquad (61)$$

$$\begin{aligned}\tilde{R}_A\left(\tilde{Z}\right) &= \tilde{R}_A^{(1)} = \tilde{R}_I^{(2)} && \tilde{Z}_A^{(0)} \le \tilde{Z} \le \tilde{Z}_A^{(1)}; \tilde{Z}_A^{(0)} = \tilde{Z}_I^{(2)} = \tilde{z}_I \Leftrightarrow A_1 \\ &= \tilde{R}_A^{(1)} + \left(\tilde{Z} - \tilde{Z}_A^{(1)}\right)\left(\tilde{R}_A^{(2)} - \tilde{R}_A^{(1)}\right)\Big/\left(\tilde{Z}_A^{(2)} - \tilde{Z}_A^{(1)}\right) && \tilde{Z}_A^{(1)} \le \tilde{Z} \le \tilde{Z}_A^{(2)} \Leftrightarrow A_2 \\ &= \tilde{R}_A^{(2)} && \tilde{Z}_A^{(2)} \le \tilde{Z} \le \tilde{Z}_A^{(3)} \Leftrightarrow A_3 \\ &= \tilde{R}_A^{(2)}\left(\tilde{Z}_A^{(4)} - \tilde{Z}\right)\Big/\left(\tilde{Z}_A^{(4)} - \tilde{Z}_A^{(3)}\right) && \tilde{Z}_A^{(3)} \le \tilde{Z} \le \tilde{Z}_A^{(4)} \Leftrightarrow A_4\end{aligned} \qquad (62)$$

The profile is illustrated in Fig 4 along with the GV surfaces up to $\tau_p$ .

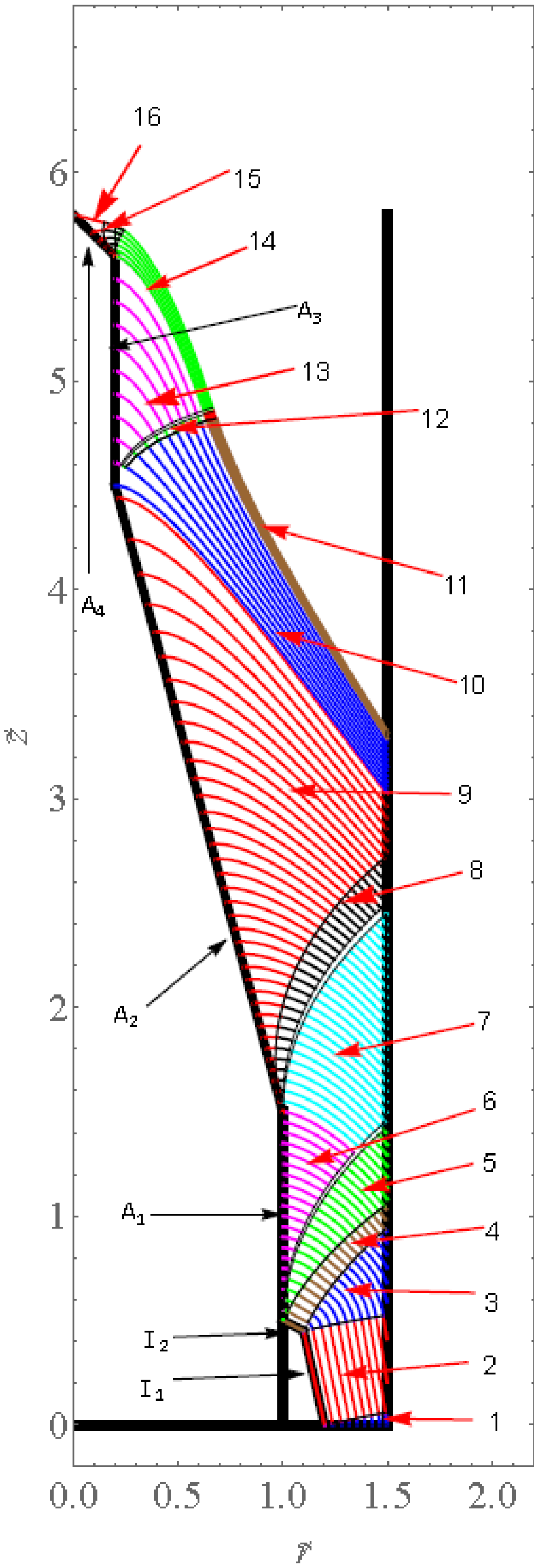

Fig 4: Device profile described by (62) shown in thick black straight lines. The parameters are as follows

$\tilde{R}_I^{(0)} = 1.2$, $\tilde{R}_I^{(1)} = 1.1$, $\tilde{Z}_I^{(1)} = 0.45$, $\tilde{Z}_A^{(0)} = \tilde{Z}_I^{(2)} = \tilde{z}_I = 0.5$ $\tilde{R}_A^{(1)} = \tilde{R}_I^{(2)} = 1;$ $\tilde{Z}_A^{(1)} = 1.5$, $\tilde{R}_A^{(2)} = 0.2$, $\tilde{Z}_A^{(2)} = 4.5$, $\tilde{Z}_A^{(3)} = 5.6$, $\tilde{Z}_A^{(4)} = 5.8$ $\tilde{r}_C = 1.5$ . This gives $\tau_1 = 2.0$, $\tau_2 = 5.7258$, $\tau_3 = 6.1658$, $\tau_4 = 6.22237$. The GV surfaces are shown as functions of $\tau$, in intervals of $\Delta\tau = 0.1$ up to $\tau_2$, in intervals of $\Delta\tau = (\tau_3 - \tau_2)/10 = 0.044$ in anode segment $A_3$ and $\Delta\tau = (\tau_4 - \tau_3)/5 = 0.011314$ in anode segment $A_4$. They are composed of 16 sub-regions, labeled with numerals and shown with different colors (in online version). Table I summarizes the parameters that are used to calculate them. The $(\tilde{r}, \tilde{z})$ values are matched at the boundaries of the sub-regions and at the cathode.

If the vertex $\tilde{Z}_I^{(1)}$ on the insulator profile is moved down to lie on the line joining the vertex at $\tilde{Z} = 0$ to that at $\tilde{Z} = \tilde{z}_I$, the two segments of insulator profile merge so that sub-region 3 of Fig 3 vanishes. If it is moved further down, the two segments again become distinct but their characteristics intersect violating the uniqueness condition. The non-existence of a classical solution may manifest itself in a discharge that proceeds from the foot of the first insulator segment to the top of the second insulator segment bypassing the intermediate vertex $\tilde{Z}_I^{(1)}$.

Contrast this with the fact that characteristics drawn from different segments of anode profile do intersect but since they become "genuine" characteristics at different times, this merely implies that the characteristics from earlier segments influence the GV surface at a time much later than the time the intersection of GV surface with the anode crosses the segment boundary.

Table I: Parameters for calculating GV surfaces in Fig 3.

| # | Segment /Vertex | $\tilde{Z}$ | N | $\tau$ | s |
|---|---|---|---|---|---|
| 1 | $\left(\tilde{R}_I^{(0)},0\right)$ | 0 | $\left(0\cdots N_I\left(\tilde{Z}_I^{(1)}\right)\right)$ | $\left(0\cdots\tau_1\right)$ | -1 |
| 2 | $I_1$ | $\left(0\cdots\tilde{Z}_I^{(1)}\right)$ | $N_I\left(\tilde{Z}\right)$ | $\left(0\cdots\tau_1\right)$ | -1 |
| 3 | $\left(\tilde{R}_I^{(1)},\tilde{Z}_I^{(1)}\right)$ | $\tilde{Z}_I^{(1)}$ | $\left(N_I^-\left(\tilde{Z}_I^{(1)}\right)\cdots N_I^+\left(\tilde{Z}_I^{(1)}\right)\right)$ | $\left(0\cdots\tau_1\right)$ | -1 |
| 4 | $I_2$ | $\left(\tilde{Z}_I^{(1)}\cdots\tilde{z}_I\right)$ | $N_I\left(\tilde{Z}\right)$ | $\left(0\cdots\tau_1\right)$ | -1 |
| 5 | $\left(\tilde{R}_A^{(1)},\tilde{z}_I\right)$ | $\tilde{z}_I$ | $\left(N_I\left(\tilde{Z}_I^{(2)}\right)\cdots N_A\left(\tilde{Z}_I^{(2)}\right)\right)$ | $\left(0\cdots\tau_2\right)$ | -1 |
| 6 | $A_1$ | $\left(\tilde{z}_I\cdots\tilde{Z}_A^{(1)}\right)$ | $N_A\left(\tilde{Z}\right)$ | $\left(0\cdots\tau_1\right)$ | 1 |
| 7 | $A_1$ extended | $\left(\tilde{z}_I\cdots\tilde{Z}(\tau)\right)$ | $N_A\left(\tilde{Z}\right)$ | $\left(\tau_1\cdots\tau_2\right)$ | 1 |
| 8 | $\left(\tilde{R}_A^{(1)},\tilde{Z}_A^{(1)}\right)$ | $\tilde{Z}_A^{(1)}$ | $\left(N_A^-\left(\tilde{Z}_A^{(1)}\right)\cdots N_A^+\left(\tilde{Z}_A^{(1)}\right)\right)$ | $\left(\tau_1\cdots\tau_2\right)$ | 1 |
| 9 | $A_2$ | $\left(\tilde{Z}_A^{(1)}\cdots\tilde{Z}(\tau)\right)$ | $N_A\left(\tilde{Z}\right)$ | $\left(\tau_1\cdots\tau_2\right)$ | 1 |
| 10 | $A_2$ extended | $\left(\tilde{Z}_A^{(1)}\cdots\tilde{Z}(\tau)\right)$ | $N_A\left(\tilde{Z}\right)$ | $\left(\tau_2\cdots\tau_3\right)$ | -1 |
| 11 | $A_2$ extended | $\left(\tilde{Z}_A^{(1)}\cdots\tilde{Z}(\tau)\right)$ | $N_A\left(\tilde{Z}\right)$ | $\left(\tau_3\cdots\tau_4\right)$ | -1 |
| 12 | $\left(\tilde{R}_A^{(2)},\tilde{Z}_A^{(2)}\right)$ | $\tilde{Z}_A^{(2)}$ | $\left(N_A^-\left(\tilde{Z}_A^{(2)}\right)\cdots N_A^+\left(\tilde{Z}_A^{(2)}\right)\right)$ | $\left(\tau_2\cdots\tau_4\right)$ | -1 |
| 13 | $A_3$ | $\left(\tilde{Z}_A^{(2)}\cdots\tilde{Z}(\tau)\right)$ | $N_A\left(\tilde{Z}\right)$ | $\left(\tau_2\cdots\tau_3\right)$ | 1 |
| 14 | $A_3$ extended | $\left(\tilde{Z}_A^{(2)}\cdots\tilde{Z}(\tau)\right)$ | $N_A\left(\tilde{Z}\right)$ | $\left(\tau_3\cdots\tau_4\right)$ | 1 |
| 15 | $\left(\tilde{R}_A^{(2)},\tilde{Z}_A^{(3)}\right)$ | $\tilde{Z}_A^{(3)}$ | $\left(N_A^-\left(\tilde{Z}_A^{(3)}\right)\cdots N_A^+\left(\tilde{Z}_A^{(3)}\right)\right)$ | $\left(\tau_3\cdots\tau_4\right)$ | 1 |
| 16 | $A_4$ | $\left(\tilde{Z}_A^{(3)}\cdots\tilde{Z}(\tau)\right)$ | $N_A\left(\tilde{Z}\right)$ | $\left(\tau_3\cdots\tau_4\right)$ | 1 |

Fig 5 demonstrates the reflection of the GV surface for a Type II anode profile. The figure shows sub-regions 14, 15 and 16 described in Table I and Fig 4 in time steps of $\Delta\tau = \left(\tau_4 - \tau_3\right)/40 = 0.00141421$ from $\tau_3$ to $\tau_4 + 0.02$.

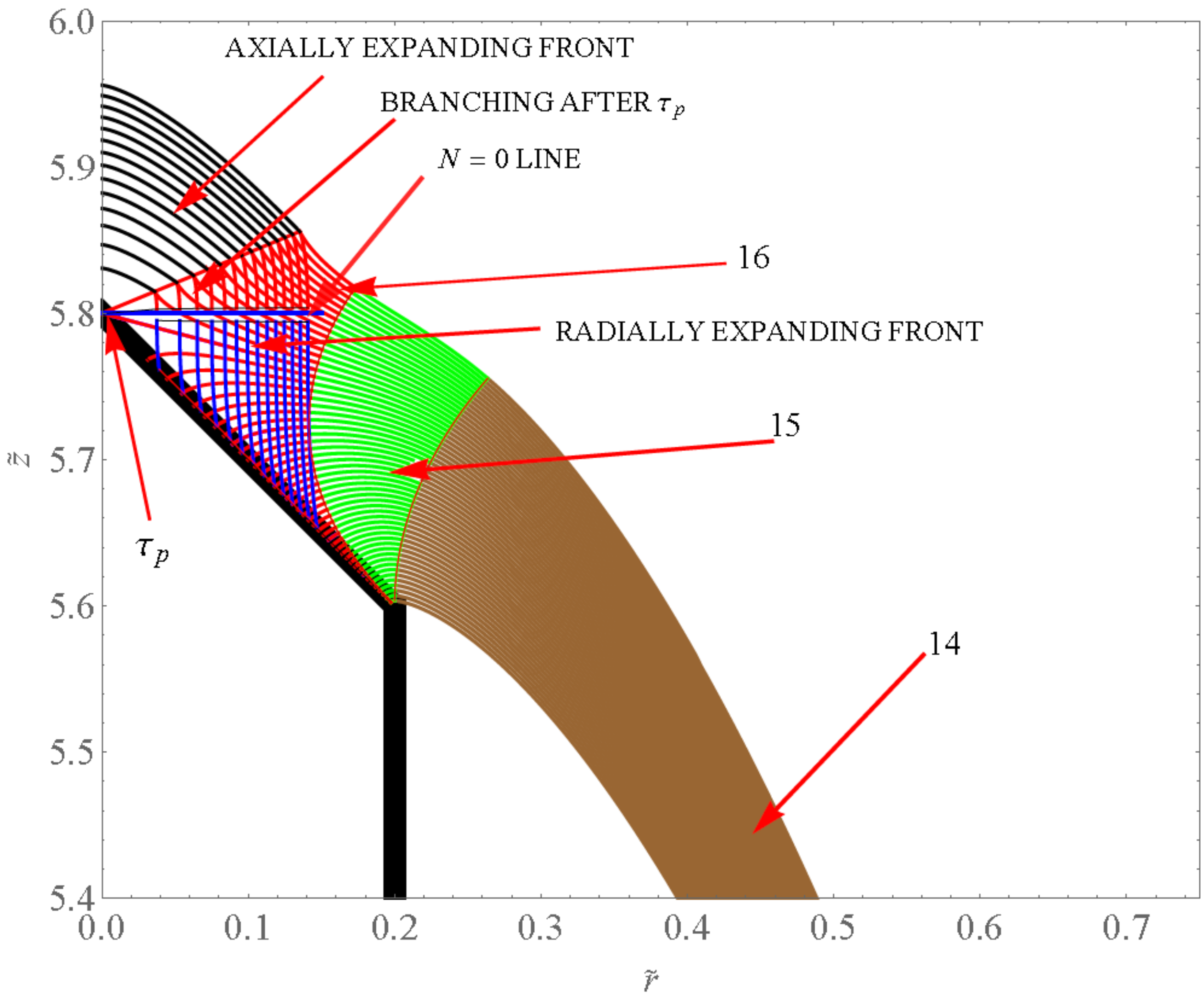

Fig 5: The GV surface in sub-region 16 travels along the tapered anode profile until it reaches the axis at $\tau_p$. For $\tau > \tau_p$, it undergoes branching. The locus of branch points is determined empirically and is shown by a red line (in online version) starting from the anode tip. The branch points are joined to the Axially Expanding Front (AEF). The lower branch of the Radially Expanding Front (REF) ends at the N=0 line. It is continued towards the anode surface by the ghost element with negative sign of N in (28).

The noteworthy aspect of the figure is the transition of the GV surface, after it reaches the anode tip at $\tau_p$, into a qualitatively different curve with two branches with the same numerical procedure. The lower branch terminates on the N=0 line and the ghost segment of anode profile with negative sign of N in equation (28) connects it with the tapered anode. The space between the locus of branch points (connected by a line emanating from the anode tip) and the axis is filled by the Axially Expanding Front described by (57).

## IV. Circuit model and closure

The time $\tau$ is proportional to the charge that has flown in time t, which has to be found by solving the circuit equation. For this purpose, the inductance of the plasma needs to be calculated using the shape of the GV surface determined above. The inductance is defined using magnetic energy of the azimuthal magnetic field:

$$\tfrac{1}{2}LI^2 \equiv \int d^3\vec{r}\,\frac{B^2}{2\mu_0} = \iint 2\pi r dr dz\left(\frac{1}{2\mu_0}\left(\frac{\mu_0 I}{2\pi r}\right)^2\right) \tag{63}$$

where the integral is over the entire volume occupied by the azimuthal magnetic field, i.e. volume enclosed between GV surface and the electrodes. Since this region varies with time, the plasma inductance can be expressed in dimensionless variables

$$\begin{aligned} L_P &= \frac{\mu_0 a}{2\pi}\iint d\tilde{z}d\tilde{r}\,\frac{1}{\tilde{r}} \equiv \frac{\mu_0 a}{2\pi}\mathfrak{L}(\tau) \\ \mathfrak{L}(\tau) &= \int_{\tilde{r}_{min}}^{\tilde{r}_C} \tilde{z}_{GV}(\tilde{r},\tau)\,\tilde{r}^{-1}d\tilde{r} \end{aligned} \tag{64}$$

The circuit equation for a DPF driven by a capacitor bank of capacitance $C_0$, internal inductance $L_0$ and internal resistance $R_0$ charged to voltage $V_0$ is

$$\frac{d}{dt}(LI) = V_0 - \frac{1}{C_0}\int_0^t I(t')dt' - IR_0 \tag{65}$$

Define

$$\begin{aligned} &\varepsilon \equiv Q_m/C_0V_0\,;\ \ \kappa \equiv \mu_0 a/2\pi L_0\,;\ I_0 \equiv V_0\sqrt{C_0/L_0}\,;\ \tilde{I}(\tau) \equiv I(\tau(t))/I_0\,; \\ &\Phi \equiv (LI)/L_0I_0 = (1+\kappa\mathfrak{L}(\tau))\tilde{I}\,;\ \gamma \equiv R_0\sqrt{C_0/L_0} \end{aligned} \tag{66}$$

Rewriting equation (10) as,

$$Q_m\frac{d\tau}{dt} = I(t);\frac{d}{dt} = \frac{d\tau}{dt}\frac{d}{d\tau} \tag{67}$$

equation (65) can be transformed as

$$\Phi \frac{d\Phi}{d\tau} = \varepsilon\left(1+\kappa\mathfrak{L}(\tau)\right)\left(1-\varepsilon\tau\right) - \varepsilon\gamma\Phi \tag{68}$$

where $\Phi = \left(1+\kappa\mathfrak{L}(\tau)\right)\tilde{I}(\tau)$ is the dimensionless magnetic flux.

The solution of this GV circuit equation is obtained by a successive approximation method [4]. The flux function $\Phi(\tau)$ is treated as the limit of a sequence of functions $\Phi_n(\tau)$, $n = 0,1,2\cdots$ obeying the equation

$$\begin{aligned} &\Phi_{n+1}\frac{d\Phi_{n+1}}{d\tau} = \varepsilon\left(1+\mathfrak{L}(\tau)\kappa\right)\left(1-\varepsilon\tau\right) - \varepsilon\gamma\Phi_n \\ &\Rightarrow \Phi_{n+1}{}^2(\tau) = \Phi_0{}^2(\tau) - 2\varepsilon\gamma\int_0^\tau d\tau\Phi_n \end{aligned} \tag{69}$$

To the zeroth order in the small parameter $\varepsilon\gamma$, it is given by the expression

$$\begin{aligned} &\Phi_0{}^2(\tau) = 2\varepsilon\tau - \varepsilon^2\tau^2 + 2\varepsilon\kappa m_0(\tau) - 2\kappa\varepsilon^2 m_1(\tau) \\ &m_0(\tau) \equiv \int_0^\tau d\tau'\mathfrak{L}(\tau');\, m_1(\tau) \equiv \int_0^\tau d\tau'\tau'\mathfrak{L}(\tau') \end{aligned} \tag{70}$$

The real time t corresponding to the independent variable $\tau$ is determined in units of the short-circuit quarter-cycle time $T_{1/4} \equiv \pi/2\cdot\sqrt{C_0 L_0}$ :

$$\tilde{t} \equiv t/T_{1/4} = \left(2\,\varepsilon/\pi\right)\cdot\int_0^\tau d\tau'/\tilde{I}(\tau') = \left(2\,\varepsilon/\pi\right)\cdot\int_0^\tau d\tau'\left(1+\kappa\mathfrak{L}(\tau)\right)/\Phi(\tau') \tag{71}$$

The fractions of stored energy converted into magnetic energy, $\eta_m$, resistively dissipated, $\eta_R$, remaining in capacitor bank, $\eta_C$, and electromagnetic work done, $\eta_w$, are given by [4]

$$\eta_m = \Phi(\tau)^2/\left(1+\kappa\mathfrak{L}(\tau)\right) \tag{72}$$

$$\eta_R = 2\varepsilon\gamma\int d\tau\tilde{I} \tag{73}$$

$$\eta_C = \left(1-\varepsilon\tau\right)^2 \tag{74}$$

$$\eta_W = 1-\left(\eta_m+\eta_C+\eta_R\right) \tag{75}$$

The condition for maximum energy transfer from the capacitor bank to the plasma at the time the plasma reaches the end of segment $\tilde{Z}_A^{(i)}$ is therefore found from (74):

$$\eta_C = \left(1 - \varepsilon\tau_i\right)^2 = 0 \qquad (76)$$

To the zeroth order of the small parameter $\varepsilon\gamma$, the magnetic energy transferred up to this time is

$$\eta_m = \left(1 + 2\tau_i^{-1}\kappa m_0\left(\tau_i\right) - 2\kappa\tau_i^{-2} m_1\left(\tau_i\right)\right) \Big/ \left(1 + \kappa\mathfrak{L}\left(\tau_i\right)\right) \qquad (77)$$

By evaluating the inductance as a function of $\tau$ for the range of interest of the profile parameters, a regression model can be established for $\mathfrak{L}\left(\tau\right)$ from which the parameters which maximize $\eta_m$ while keeping $\eta_C = 0$ can be determined.

## V. Summary and conclusion

This paper is an abridged version of a larger project aimed at constructing an incremental model of the plasma focus and similar devices. This version is focused on the energy transfer optimization problem for a plasma focus with arbitrary anode and insulator shapes. It is likely to be further revised with additional illustrative examples.

23. The solution to equation (46) might at first glance appear to be the trivial solution $\alpha = \beta$. The procedure used to obtain a non-trivial solution is (1) assign a numerical value to $\left(N/\tilde{r}_v\right)$ (2) Substitute $\beta = 2\mathrm{ArcSech}\left(N/\tilde{r}_v\right)$ in LHS of (46) and numerically find its roots $\alpha_{sol}$ (3) Tabulate values of $\left(N/\tilde{r}_v\right)$ and $\alpha_{sol}$. It is verified that besides the trivial solution, there is another root that leads to the envelope curve.